\documentclass[graybox]{svmult}

\usepackage{hyperref}
\usepackage{type1cm}        
\usepackage{makeidx}         
\usepackage{graphicx}        
\usepackage{multicol}        
\usepackage[bottom]{footmisc}

\usepackage{ulem}
\usepackage{newtxtext}       %
\usepackage[varvw]{newtxmath}       

\makeindex             

\begin{document}

\title*{Random resetting in search
problems}
\author{Arnab Pal, Viktor Stojkoski and Trifce Sandev}
\institute{Arnab Pal \at  The Institute of Mathematical Sciences, Chennai, 600113, India \& Homi Bhabha National Institute, Mumbai, 400094, India. \email{arnabpal@imsc.res.in}
\and Viktor Stojkoski \at  Faculty of Economics, Ss. Cyril and Methodius University, Skopje, 1000, Macedonia. \email{vstojkoski@eccf.ukim.edu.mk}
\and Trifce Sandev \at Macedonian Academy of Sciences and Arts, Skopje, 1000, Macedonia \& Institute of Physics, Faculty of Natural Sciences and Mathematics, Ss. Cyril and Methodius University, Skopje, 1000, Macedonia. \email{trifce.sandev@manu.edu.mk}}
%
%
\maketitle


\abstract*{Each chapter should be preceded by an abstract (no more than 200 words) that summarizes the content. The abstract will appear \textit{online} at \url{www.SpringerLink.com} and be available with unrestricted access. This allows unregistered users to read the abstract as a teaser for the complete chapter.
Please use the 'starred' version of the \texttt{abstract} command for typesetting the text of the online abstracts (cf. source file of this chapter template \texttt{abstract}) and include them with the source files of your manuscript. Use the plain \texttt{abstract} command if the abstract is also to appear in the printed version of the book.}

\abstract{By periodically returning a search process to a known or random state, random resetting possesses the potential to unveil new trajectories, sidestep potential obstacles, and consequently enhance the efficiency of locating desired targets. In this chapter, we highlight the pivotal theoretical contributions that have enriched our understanding of random resetting within an abundance of stochastic processes, ranging from standard diffusion to its fractional counterpart. We also touch upon the general criteria required for resetting to improve the search process, particularly when distribution describing the time needed to reach the target is broader compared to a normal one. Building on this foundation, we delve into real-world applications where resetting optimizes the efficiency of reaching the desired outcome, spanning topics from home range search, ion transport to the intricate dynamics of income. Conclusively, the results presented in this chapter offer a cohesive perspective on the multifaceted influence of random resetting across diverse fields.}




\section{Introduction}
\textit{When in doubt, reset! }

Random resetting is a concept that has gained substantial attention in recent years as a novel and counter-intuitive approach to solving complex search problems. This technique has provided groundbreaking insights into the stochastic nature of search processes whose behavior can often confound traditional models and approaches~\cite{evans2020stochastic}. In this chapter, we delve into the fascinating interplay between random resetting and search processes. We explore how resetting randomly to a new location \cite{evans2020stochastic,evans2011diffusion,pal2015diffusion,evans2013optimal,eule2016non,nagar2016diffusion,mendez2016characterization,kumar2023universal,kusmierz2014first,bonomo2021first,belan2018restart} acts as a mechanism that provides new insights into the dynamics of various stochastic systems, enhancing our ability to predict and control their behavior.

\textit{But why would random resetting improve the modeling of search processes?}\\

\begin{figure}[t]
    \centering
    \includegraphics[scale=0.55]{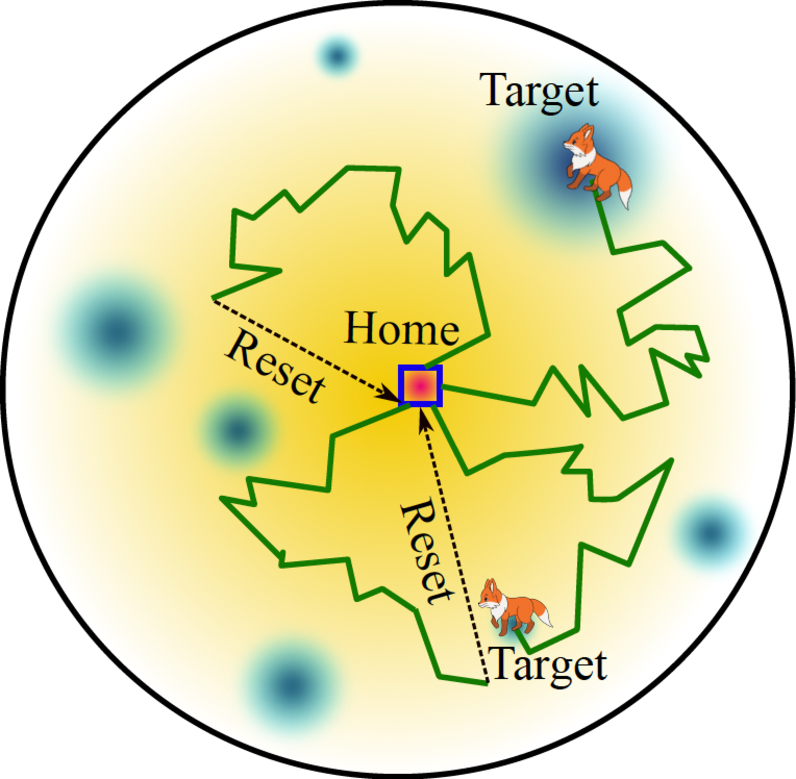}
    \caption{Schematic for a stochastic search process under resetting in a confined topography in the presence of multiple targets (dark green circles). The searcher (e.g., a prairie fox) looks for targets (which can be food, companion or other resources) in the vicinity of its home and intermittently returns there to take rest. Following this \textit{spatial reset}, the searcher resumes its search, more likely wandering to a different direction. The search is completed after it has found one or multiple targets in the domain. The central aim here is to understand the behavior of the mean search time under random resetting, in particular, as a function of the frequency of resetting or return.}
    \label{Fig1}
\end{figure}

By periodically returning the search process to a known or random state, random resetting uncovers new pathways, avoids potential pitfalls, and can lead to an unexpected efficiency in reaching desired targets~\cite{evans2011diffusion,evans2011diffusionJPA,evans2020stochastic,pal2017first,reuveni2016optimal,chechkin2018random}. Consider home range search, a realm where the influence of random resetting has yielded unprecedented advances in efficiency and understanding~\cite{pal2020search,tal2020experimental,besga2020optimal}. In ecological studies, animals often search within a specific home range to find food, mates, or shelter. Traditional models may struggle to accurately represent the stochastic nature of this behavior. The introduction of random resetting, where the search is periodically returned to the ``home'' of the searcher or a random state, helps model the realistic unpredictability of search patterns within an animal's home range (see Fig. \ref{Fig1}). This approach not only simulates the dynamic interplay between an animal's instinctive behavior and random environmental factors, leading to more accurate representations, but also reveals that search with home returns can outperform free-range search, particularly in conditions of high uncertainty~\cite{pal2020search}.

Now consider the field of economics, specifically in the modeling of individual income dynamics, where random resetting introduces a vital perspective~\cite{stojkoski2022income,gabaix2016dynamics,vinod2022nonergodicity}. Understanding the time needed for an individual to improve their income is crucial for various economic decisions, from personal financial planning to governmental policy-making~\cite{jolakoski2023first}. Traditional models, however, may overlook the complex and often abrupt changes in an individual's financial situation, such as losing a job or retirement. By incorporating random resetting, which can mimic these sudden transitions, economists are provided with a more realistic portrayal of the time dynamics involved in income improvement. This approach not only offers a nuanced depiction of income trajectories but also recognizes the stabilizing effect of these ``reset'' points, acknowledging their real-world relevance. 

But the application of random resetting in search problems goes way beyond these two examples. From the intricate pathways of cellular biology~\cite{boyer2014random,roldan2016stochastic,ramoso2020stochastic,budnar2019anillin}, chemical reactions \cite{reuveni2014role,rotbart2015michaelis,biswas2023rate,budnar2019anillin}, foraging \cite{pal2020search,moran2023modeling,benichou2011intermittent}, nonlinear dynamical systems \cite{ray2021mitigating}, operation research~\cite{bonomo2022mitigating} to computer algorithms \cite{luby1993optimal,blumer2022stochastic}, the principles of random resetting continue to offer fresh insights and innovative solutions. The versatility and adaptability of this concept have allowed researchers and practitioners to tackle previously intractable challenges, forging connections between disparate fields and creating a cohesive understanding of the underlying dynamics of search processes. 

In this chapter, we systematically explore the multifaceted realm of search processes with random resetting. In Section~\ref{sec:formalism}, we introduce a robust mathematical framework, elucidating the core concepts and recent advancements that have shaped our understanding of search under the influence of random resetting. Moving forward, in Section~\ref{sec:examples-r}, we delve into examples of search processes under resetting, offering a detailed examination of the intriguing theoretical behaviors and patterns that arise from the interplay between search strategies and random resets. This includes a focus on how random resetting can influence metrics such as first passage time. Next, in Section~\ref{sec:when-resetting-works}, we explore the delicate balance between too much and too little resetting, investigating strategies that maximize efficiency in search processes by optimizing the resetting rate. Then, in Section~\ref{sec:applications}, we showcase diverse applications of search processes under random resetting, demonstrating the wide-ranging impact of this approach on contemporary science and technology, as well as its practical solutions to real-world challenges. Finally, 
we synthesize the insights gleaned from our exploration, highlighting both the theoretical underpinnings and practical implications of incorporating random resetting into search processes.

\section{The renewal formalism for first passage  under resetting}
\label{sec:formalism}
In this section, we discuss the first passage properties of a resetting process. Let us consider a generic search process conducted by a stochastic searcher $\Vec{x}(t)$ in an arbitrary domain $\mathcal{D}$ in the presence of a single or multiple targets. The initial condition of the process is assumed to be $\Vec{x}(0)=\Vec{x_0}$.  The process is completed when the searcher finds one of these targets and one would be interested in the statistics of the stochastic search time $T$, distribution of which is denoted by the first passage time density   $f_T(t|\Vec{x_0})$. It is often useful to introduce $Q_0(t|\Vec{x_0})$ that defines the survival probability of the underlying process that it has not found any target until time $t$, starting from the initial configuration $\Vec{x_0}$. Clearly, this quantity is related to the position distribution function, denoted by $P_0(\Vec{x},t|\Vec{x_0},0)$, of the searcher 
in the following way $Q_0(t|\Vec{x_0})=\int_{\mathcal{D}}~d\Vec{x}P_0(\Vec{x},t|\Vec{x_0},0)$. In the absence of targets, the searcher always survives thus $Q_0(t|\Vec{x_0})=1$, otherwise it decays to zero in large time as the searcher eventually finds the target. In other words, the search process is completed.

To introduce resetting, let us consider that the underlying first passage process is intermittently stopped and restarted from some pre-selected configuration. Simply put, after each resetting event, the searcher goes back to a fixed location $\Vec{x_R}$ or locations drawn from an identical ensemble $P(\Vec{x_R})$. The waiting times between the resetting events are drawn from a normalized density $f_R(t)$. Under this mechanism, the new first passage time is denoted by $T_r$ which will be distributed via the first passage time density $f_{T_r}(t|\Vec{x_0})$. The subscript `$r$' is used to indicate the presence of resetting events. Here, $Q_r(t|\Vec{x_0};\Vec{x_R})$ denotes the survival probability in the presence of resetting. In words, this measures the probability to find the searcher inside the domain $\mathcal{D}$ upto time $t$ given that it had started at $\Vec{x_0}$ at time zero and experienced multiple resetting to $\Vec{x_R}$. In this chapter, we assume that the waiting times between resetting are drawn from $f_R(t)=re^{-rt}$ which essentially means that resetting occurs at a rate $r$.

Since each resetting event compels the searcher to start from scratch, the first passage process under resetting generically belongs to the broad class of stochastic renewal processes. Adapting the mathematical structure from there, one can write a renewal equation for the survival probability $Q_r(t|\Vec{x_0};\Vec{x_R})$ in the following way \cite{evans2011diffusion,evans2020stochastic}
\begin{align}
Q_r(t|\Vec{x_0};\Vec{x_R})=e^{-rt}Q_0(t|\Vec{x_0})+r\int_0^t d\tau_l e^{-r\tau_l} Q_0(\tau_l|\Vec{x_R})Q_r(t-\tau_l|\Vec{x_0};\Vec{x_R})~.
\label{Qr-renewal}
\end{align}
Eq. \ref{Qr-renewal} has a simple interpretation. The first term on the right hand side implies that the particle survives till time $t$ without experiencing any reset event. The second term considers the possibility when there are multiple reset events. One can then look at a long trajectory where the last reset event had occurred at time $t-\tau_l$, and after that there has been no reset for the duration $\tau_l$. This probability is given by $rd\tau_l e^{-r\tau_l}$. But then this has to be multiplied by $Q_r(t-\tau_l|\Vec{x_0};\Vec{x_R})$, i.e., the probability that the particle survives till time $t-\tau_l$ with multiple reset events and $Q_0(\tau_l|\Vec{x_R})$, i.e., the survival probability of the particle for the last non-resetting interval $\tau_l$, starting from $\Vec{x_R}$. This is a useful formula since the survival properties for the resetting process can be directly understood from the reset-free processes. Notably, the construction of the above equation relied upon the last resetting event, hence this approach is often known as the \textit{last renewal formalism}. Similar to this, a \textit{first renewal formalism} can also be developed for the survival probability. The corresponding renewal equation reads \cite{pal2016diffusion}
\begin{align}
Q_r(t|\Vec{x_0};\Vec{x_R})=
e^{-rt}Q_0(t|\Vec{x_0})+r\int_0^t d\tau_f e^{-r\tau_f} Q_0(\tau_f|\Vec{x_0})Q_r(t-\tau_f|\Vec{x_R};\Vec{x_R})~, 
\label{Qr-first-renewal}
\end{align}
where we assume that the first resetting occurs at time $\tau_f$, and upto that there was no resetting (with the probability $e^{-r\tau_f}$). Until then the particle survives with $Q_0(\tau_f|\Vec{x_0})$ -- the reset free survival probability. For the rest time interval $t-\tau_f$, the particle survives with many resetting events -- the probability of which is assigned by $Q_r(t-\tau_f|\Vec{x_R};\Vec{x_R})$. It is easy to see that both (\ref{Qr-renewal}) and (\ref{Qr-first-renewal}) are identical (see \cite{evans2020stochastic,pal2016diffusion}). Notably, the renewal formalism does not require any particular choice of the underlying dynamics i.e., the renewal equations hold both for the Markovian and the non-Markovian anomalous search processes. The only key assumption here is that no memory of the dynamics can be carried forward from one interval to the next. Crucially, as we show below, that it is not required to have the expression for the survival probability $Q_0(t|\cdot)$ of the underlying process in the real time domain - remarkably, an expression in the Laplace space is sufficient to extract further information. To see this, let us apply the Laplace transform on the both sides of Eq. (\ref{Qr-renewal})
which satisfies
\begin{align}
\widetilde{Q_r}(s|\Vec{x_0};\Vec{x_R})=\frac{\widetilde{Q_0}(s+r|\Vec{x_0})}{1-r \widetilde{Q_0}(s+r|\Vec{x_R})}~,
\label{renewal-qr}    
\end{align}
where $\widetilde{\mathcal{G}}(s)=\mathscr{L}_{t \to s}\left[\mathcal{G}(t)\right]$ is the Laplace transform of the function $\mathcal{G}(t)$. The first passage time density $f_{T_r}(t|\Vec{x_0})$ is the negative gradient of the survival probability in time such that \cite{redner2001guide,bray2013persistence}
\begin{align}
    f_{T_r}(t|\Vec{x_0};\Vec{x_R})=-\frac{\partial Q_r(t|\Vec{x_0};\Vec{x_R})}{\partial t}
\end{align}
which essentially provides the statistics of the first passage time under resetting. In Laplace space, this relation translates to 
\begin{align}
\widetilde{f_{T_r}}(s|\Vec{x_0};\Vec{x_R})=1-s \widetilde{Q_r}(s|\Vec{x_0};\Vec{x_R}),
\label{lt-prop-r}    
\end{align}
where we have assumed that the following boundary conditions in time $Q_r(t=0|\Vec{x_0};\Vec{x_R})=1,~Q_r(t \to \infty|\Vec{x_0};\Vec{x_R})<\infty$. The resulting relation (\ref{lt-prop-r}) is quite useful since the moments can be computed readily such as
\begin{align}
\label{relation1}
\langle T_r^{n}(\Vec{x_0};\Vec{x_R}) \rangle=(-1)^n~\frac{\mathrm{d}^n}{\mathrm{d}s^n}\widetilde{f_{T_r}}(s|\Vec{x_0};\Vec{x_R})\bigg|_{s \to 0}    
\end{align}
For instance, the mean first passage time (MFPT) under resetting reads (by taking $n=1$ in Eq. (\ref{relation1}))
\begin{align}
\langle T_r \rangle &= \frac{\widetilde{Q_0}(r|\Vec{x_0})}{1-r \widetilde{Q_0}(r|\Vec{x_R})}=\frac{1-\widetilde{f_{T}}(r|\Vec{x_0})}{r \widetilde{f_{T}}(r|\Vec{x_R})}, \label{mean under restart} 
\end{align}
where we have used two equivalent forms for the mean time under resetting -- one in terms of the survival probability and the other in terms of the first passage time density of the resetting free underlying process. Notation wise we have suppressed the dependence of the MFPT on the initial coordinate $\Vec{x_0}$ and resetting coordinate $\Vec{x_R}$ for brevity. In the remainder sections, we will mostly assume $\Vec{x_0}=\Vec{x_R}$ (unless otherwise stated) without loss of much generality.

\section{Applications to theoretical first passage models}
\label{sec:examples-r}
In this section, we review a few canonical examples of search processes under resetting mechanism. The central goal is to show how the formalism developed in the previous section can be applied to all these examples in a unified manner. In doing so, we delve deeper into the ramifications due to resetting on the first passage statistics. Furthermore, we discuss about the search optimization conditions in resetting induced processes. 

\subsection{Diffusion with stochastic resetting}
Let us consider a diffusion mediated search in one dimension where a Brownian particle diffuses through the medium starting from $x_0$ at time zero~\cite{stojkoski2022autocorrelation}. Motion of the particle can be quantified in terms of the probability density function $P(x,t)$ for the position $x$ at time $t$, which is given by the diffusion equation
\begin{align}
    \frac{\partial P(x,t)}{\partial t}=D \frac{\partial^2 P(x,t)}{\partial x^2},
\end{align}
where $D$ is the diffusion constant. The particle is also reset to $x_0$ intermittently with a rate $r$. We assume that there is an absorbing boundary at the origin and we are interested in the mean time for the particle to find the boundary. This metric can be computed by employing Eq. (\ref{mean under restart}) where we have to use the survival probability of the resetting free process. The latter is well known in the literature (see e.g. \cite{redner2001guide}) and is given by 
\begin{align}
Q_0(t|x_0)=\int_0^L~dx P(x,t)=\text{erf}\left[\frac{x_0}{\sqrt{4Dt}}\right].
\label{survival1}
\end{align}
from which one can find the following in Laplace space
\begin{align}
\label{surivival-LT}
\widetilde{Q_0}(s|x_0)=\frac{1}{s} \left[ 1-e^{-\sqrt{\frac{s}{D}}x_0} \right].
\end{align}
The first passage time density is also straightforward to compute
\begin{align}\label{FP-LT}
    \widetilde{f_T}(s|x_0)=e^{-\sqrt{\frac{s}{D}}x_0}.
\end{align}
Substituting either of the above namely Eq. (\ref{surivival-LT}) or Eq. (\ref{FP-LT}) into Eq.~(\ref{mean under restart}) gives the 
MFPT of one dimensional diffusion under resetting which was first derived by Evans and Majumdar
\cite{evans2011diffusion}
\begin{align}\label{mfpt_Br}
    \langle T_r \rangle &=\frac{1}{r} \left[ e^{\alpha x_0}-1  \right], 
\end{align}
where $\alpha=\sqrt{\frac{r}{D}}$ is an inverse length that corresponds to the typical distance covered by the particle between two resetting events. Note that $\underset{r \to 0}{\lim} \langle T_r \rangle   \sim \frac{1}{r}$ which diverges since the particle hardly experiences any resetting event, and eventually drifts away from the origin. This is no surprise as for simple 1D diffusion $\langle T \rangle=\infty$.  In the other extreme limit $r \to \infty$, the mean first passage time also diverges exponentially since the particle almost remains localized around the resetting coordinate. Evidently, this marks the existence for an optimal resetting rate $r^*$ which can be computed by setting 
\begin{align}
    \frac{d \langle T_r \rangle }{dr}\bigg|_{r=r^*}=0
    \label{MFPT-diff-opt}
\end{align}
Using the MFPT for diffusion under resetting from Eq. (\ref{mfpt_Br}) in Eq. \ref{MFPT-diff-opt} and defining $z^*=\sqrt{\frac{r^*}{D}}x_0$, we obtain the following equation for the optimal resetting rate
\begin{align}
\frac{z^{*}}{2}=1-e^{-z^* } ~.
\label{optimize-unbounded-1-diff}
\end{align}
Eq. 
\eqref{optimize-unbounded-1-diff} gives $z^{*}=1.5936...$ from which one finds $r^* \sim 2.54 \tau_{d}^{-1}$, where $\tau_{d}=x_0^2/D$ is the diffusive time scale. In terms of this diffusive time scale, the minimum MFPT becomes $ \langle T_{r*} \rangle \sim 1.544 \tau_{d}$. Thus, the minimal first passage time is obtained when resetting is conducted at a diffusive time scale.

\subsubsection{Diffusion with resetting in higher dimensions}
As mentioned above, the formalism can be applied also to a search process that is being conducted in an arbitrary spatial dimension $d$. Consider a diffusive searcher
starts at the 
initial position $\Vec{x_0}$ and undergoes stochastic resetting
to  $\Vec{x_0}$ with a constant rate $r$. There is a finite size target -- an absorbing  $d$-dimensional sphere of radius $a$  (with $|{\vec x_0}|>a$)
centred at the origin. Whenever the searcher reaches
the surface of the target sphere, the particle is absorbed. The expression for the survival probability of the diffusive particle starting from $\Vec{ x_0}$ with the absorbing
sphere at the origin, is given by the following expression in Laplace space (see e.g.~\cite{redner2001guide})
\begin{equation}
\widetilde{ Q_0}(s|\Vec x_0) =\frac{1}{s} - \frac{1}{s}  \frac{R_0^\nu}{a^\nu} \frac{ K_\nu (\sqrt{s/D} R_0)}{ K_\nu (\sqrt{s/D} a)}\;,
\end{equation}
where $K_\nu (z)$ is the modified Bessel function of the second kind, with index $\nu = 1-d/2$. Here, $R_0 = |\Vec x_0|$ is the distance from the resetting position to the target. Using Eq. (\ref{mean under restart}), one arrives at the following expression for the MFPT of the searcher to the target sphere~\cite{evans2014diffusion}
\begin{align}
    \langle T_r \rangle = \frac{1}{r} \left[ \frac{a^\nu}{R_0^\nu} \frac{K_\nu(\sqrt{r/D}a)}{K_\nu(\sqrt{r/D}R_0)}-1  \right].
\end{align}
 Similar to one dimensional diffusion, resetting will always be useful to render a finite MFPT for such $d$-dimensional diffusive search in unconfined space.

\subsection{Diffusive search in a potential landscape}
A Brownian particle diffusing in a potential landscape $U(x)$ is described by the following Smoluchowski or Fokker-Planck equation
\begin{align}
\label{fpe}
\frac{\partial}{\partial t}P(x,t)=\left(\frac{\partial}{\partial x}U'(x)+
D\frac{\partial^2}{\partial x^2}\right)P(x,t),
\end{align}
with the initial condition $P(x,0)=\delta(x-x_0)$. In what follows, we consider different potentials and study the trade-off between the resetting and attraction due to the potential.

\subsubsection{Linear potential}
Let us first consider a case where the particle starts from $x_0$ and experiences a linear potential $U(x)=k|x|$, the minimum of which is centered at the origin. In addition, the particle is reset to $x_0$ at a rate $r$ and one is interested in the mean time that it takes for the searcher to reach the origin which is also an absorbing boundary. In this case, the first passage time density for the underlying process in Laplace space reads \cite{pal2019local,singh2020resetting,ray2019peclet}
\begin{align}
    \widetilde{Q_0}(s|x_0)=\frac{1}{s}\left[ 1-e^{-\frac{x_0}{2D} \left( \sqrt{k^2+4Ds} -k \right)} \right].
\end{align}
Substituting the above in Eq. (\ref{mean under restart}), we find
\begin{align}
  \langle T_r \rangle=\frac{1}{r}\left[e^{\frac{x_0}{2D} \left( \sqrt{k^2+4Dr}-k \right)}-1\right]~,
\label{sol:mfptx_0-unbound}
\end{align}
which in the limit $r \to 0$ becomes $|x_0|/k$. In this case, the optimal resetting rate $r^*$ can be obtained by solving the following transcendental equation \cite{pal2019local}
\begin{align}
\frac{z^{*2}}{2\sqrt{Pe^2+z^{*2}}}=1-e^{- \left( \sqrt{Pe^2+z^{*2} } -Pe\right) }~,
\label{optimize-unbounded-1}
\end{align}
where recall $z^*=\sqrt{\frac{r^*}{D}}x_0$ is the rescaled resetting rate and $Pe=kx_0/2D$ is the P\'eclet number. Eq. 
(\ref{optimize-unbounded-1}) asserts that the root $z^*$ depends on the P\'eclet number \cite{redner2001guide}. While for small P\'eclet number $(Pe<1)$ , the limit is similar to the simple diffusion case, and one has $z^*>0$. This is not the case with the high P\'eclet number limit $(Pe>1)$ which results in an approximate solution $z^* \approx 0$ (which is also a trivial solution of Eq. (\ref{optimize-unbounded-1})). This means that as one varies $Pe$, the optimal resetting rate switches between a finite value to zero. More on the physical ground, in the diffusion dominated regime resetting remains beneficial (resulting in $r^*>0$) while in the high force gradient limit the searcher is able to find the target quite efficiently making resetting only detrimental ($r^*=0$) to the search.

\subsubsection{Harmonic potential}
Let us now consider a Brownian particle in a harmonic potential $U(x)=\frac{1}{2}kx^2$ that starts from $x_0$ and tries to find the target which is located at the origin in the presence of resetting to $x_0$~\cite{pal2015diffusion}. 
Starting again from the Laplace space backward Fokker-Planck equation with appropriate boundary conditions, one finds \cite{gupta2020stochastic,ahmad2019first}
\begin{align}
\widetilde{Q}(s|x_0)=\frac{1}{s} \left[ 1-\frac{\Gamma \left(\frac{1}{2}+\frac{s}{2k}\right)}{\sqrt{2\pi}} x_0 \sqrt{k/D}~\mathcal{U}\left(\frac{1+s/k}{2},\frac{3}{2},\frac{k  x_0^2}{2D}\right)  \right],
\end{align}
where $\mathcal{U}(a,b,z)$ is the Tricomi confluent hypergeometric function \cite{zaitsev2002handbook}. The MFPT in this case reads
\begin{align}
    \langle T_r \rangle = \frac{1}{r} \left[ \frac{\sqrt{2 \pi}}{\Gamma \left(\frac{1}{2}+\frac{r}{2k}\right) x_0 \sqrt{k/D}~\mathcal{U}\left(\frac{1+r/k}{2},\frac{3}{2},\frac{k  x_0^2}{2D}\right)}-1 \right].
\end{align}
The analysis for the optimal resetting rate is similar to the discussion in the previous section. Further studies related to search processes under resetting mediated by logarithmic and power law potential were studied in \cite{ray2020diffusion} and \cite{capala2023optimization} respectively.

\subsection{Search in confined geometry}
Many search processes are often conducted in a confined domain. In such cases, the mean search time usually remains finite and it is not apparent whether resetting strategy can be of any use. In this section, we aim to gain such insights by reviewing some canonical stochastic search processes.

\subsubsection{Diffusive search in a confinement under resetting}
\label{1D-interval}
Consider a Brownian particle, initially located at $x_0$,
diffusing in an interval $[0, L]$ in one dimension. The particle can get absorbed by any of these boundaries. In addition,
the particle is stochastically reset to the initial position $x_0$ and we are interested in the first-passage
properties of the particle due to resetting. The probability density $P(x,t)$ of the underlying process is a classical result and known from the literature~\cite{redner2001guide}
\begin{align}
    P(x,t)=\dfrac{2}{L}\sum_{n=1}^{\infty}  \psi_n(x_0)\psi_n(x) e^{-k_n t}~.
\label{propagator-no-reset}
\end{align}
where $\psi_n(x)=\sin[n \pi x/L]$ are the eigenfunctions and $k_n=n^2 \pi^2D/L^2$ is the rate at which the $n$-th eigenmode $\psi_n(x)$ decays with time. From this one can easily compute the survival probability in Laplace space \cite{pal2019firstV,pal2019landau}
\begin{align}
    \widetilde{Q_0}(s|x_0)=\frac{1}{s} \left[ 1-g_0(x_0,s) \right],\text{with~} g_0(x_0,s)=\frac{\sinh[(L-x_0) \sqrt{s/D}]+\sinh[ x_0 \sqrt{s/D}]}{\sinh [L\sqrt{s/D}]},
\label{BFP-qr}
\end{align}
The MFPT for this case then can be found substituting the survival probability into Eq. (\ref{mean under restart})
\begin{align}
     \langle  T_r \rangle=\frac{L^2}{4D}\mathcal{G}(\beta,u)~, \text{where~~}\mathcal{G}(\beta,u)=\frac{1}{\beta^2}\left[\frac{\cosh(\beta)}{\cosh \beta(1-2u)}-1\right]~,
 \label{scaling}
\end{align}
where $u=x_0/L,~\beta=\frac{L}{2}\alpha$. To find the optimal restart rate, we scale $r^*=4D{\beta^*}^2/b^2$, in terms of $\beta^*$. In this case, one can determine the domain in which resetting expedites the completion of the underlying process: $\mathscr{D}=[ (0, u_{-}) ~\cup~ (u_{+},1) ]$, where $u_{\pm}=(5\pm \sqrt{5})/10$.  When $u_{-}<u<u_{+}$, the function is minimum at $\beta^*=0$, meaning $\langle T_r \rangle$ can not be made lower by introducing resetting. On the other hand, when $u \in \mathscr{D}$, one can minimize the scaling function $\mathcal{G}(\beta,u)$ at a finite $\beta$ implying that the mean first passage time is optimized at a finite resetting rate \cite{pal2019firstV}. Similar analysis can also be extended for diffusive search in the presence of drift \cite{pal2019landau} and generically in the presence of a potential \cite{ahmad2022first} in an one dimensional confinement.

One can also consider $d$-dimensional search where a  Brownian searcher diffuses between two d-dimensional concentric spheres, starting from a distance $|\Vec{x_0}|$ to the center of sphere, $R_1 < |\Vec{x_0}| < R_2$
with $R_{1(2)}$ being the radius of the inner (outer) sphere. The process is completed once the searcher hits the inner or outer spherical surface. In addition, the searcher is reset to $|\Vec{x_0}|$ at a rate $r$. The first passage properties in the presence of resetting were discussed in \cite{chen2022first}. Several extensions were made for the similar set-up in the presence of external potential \cite{ahmad2020role,ahmad2023comparing}. First passage properties under resetting in a two dimensional circle were studied in \cite{chatterjee2018diffusion}. Escape properties of an underdamped particle with resetting from an interval were studied in \cite{capala2021random}.

\subsubsection{Anomalous processes in a confinement under resetting}
We consider a particle, initially located at $x_0$, performing a random walk in continuous time within an interval $[0,L]$
in one dimension in the presence of resetting. The walker can get absorbed by any of
these boundaries and one is interested in finding the first-passage time of the walker. We further assume that the walker is a generalized continuous time random walker \cite{metzler2000random,klafter2011first,mendez2021continuous} such that its characteristic jump distances are small compared to the interval length but the waiting times between the jumps are taken from a distribution $\phi(t)$. The master equation for this process can be written as \cite{mendez2022nonstandard}
\begin{equation}
\frac{\partial P(x,t)}{\partial t}=\frac{\overline{\sigma}^2}{2}\int_{0}^{t}K(t-t')\frac{\partial^2 P(x,t')}{\partial x^2}dt'.\label{eq:me}
\end{equation}
where $K(t)$ is the memory kernel which is related to the waiting time density $\phi(t)$. These two quantities are related to each other in Laplace space such that
\begin{equation}
\widetilde{K}(s)=\frac{s\widetilde{\varphi}(s)}{1-\widetilde{\varphi}(s)},\label{eq:mk}
\end{equation}
where $\widetilde{\varphi}(s)$ is the LT of the waiting time density $\phi(t)$. 
Applying suitable boundary conditions, one can obtain the following expression for the survival probability in the Laplace space
\begin{align}
\widetilde{Q_{0}}(s|x_{0})=\frac{1}{s}\left[1-\frac{\cosh\left(\overline{\alpha} (s)\left( x_{0}-\frac{ L}{2}\right)\right)}{\cosh\left(\overline{\alpha} (s)\frac{L}{2}\right)}\right],
\label{s0}
\end{align}
where $\overline{\alpha}(s) \equiv \frac{1}{\overline{\sigma}}\sqrt{\frac{2s}{\widetilde{K}(s)}}=\frac{\sqrt{2}}{\overline{\sigma}}\sqrt{\frac{1}{\widetilde{\varphi}(s)}-1}$. 
Substituting the above in Eq. (\ref{mean under restart}), the MFPT under resetting reads \cite{mendez2022nonstandard}
\begin{align}
\langle T_r \rangle
=\frac{1}{r}\left[\frac{\cosh\left(\frac{\overline{\alpha} (r)L}{2}\right)}{\cosh\left(\overline{\alpha} (r)\left(x_{0}-\frac{L}{2}\right)\right)}-1\right],
\label{tr2}
\end{align}
which holds for arbitrary memory kernel. One can study the MFPT under quite some generalities such as waiting time with finite first and second moments, with finite first moment and diverging second moment and diverging first and second moments. In the former case, one can derive a condition on $x_0/L$ for resetting to be beneficial (this is similar to the diffusion as was discussed in Sec. \ref{1D-interval}) while in the latter two cases, resetting always expedites the completion marking the existence of a finite optimal resetting rate \cite{mendez2022nonstandard}.

\subsection{Fractional diffusion with stochastic resetting}
Consider a subdiffusion process that undergoes stochastic resetting. The set-up is similar to before -- the walker starts at $x_0>0$ at time zero and the process is completed when it finds the boundary at the origin. The corresponding Fokker-Planck equation for the fractional diffusion process is given by~\cite{sokolov2005diffusion
}
\begin{align}
    \frac{\partial P(x,t)}{\partial t}=D\frac{d}{dt}\int_{0}^{t}\eta(t-t')\frac{\partial^2 P(x,t')}{\partial x^2}dt'.
\end{align}
where the memory kernel $\eta(t)$ emanates from the waiting time distribution $\phi(t)$ of the walker between the jumps. This relation, in Laplace space, reads $\widetilde{\phi}(s)=\frac{1}{1+1/\widetilde{\eta}(s)}$. Note that Eq. (\ref{eq:me}) is related to the above equation with the transformation $\widetilde{\eta}(s)\rightarrow \widetilde{K}(s)/s$. For this process, the survival probability of a particle starting at $x_0$ at time zero in the absence of resetting reads \cite{stanislavsky2021optimal}
\begin{equation}
    \widetilde{Q_0}(s|x_0) =  
    \frac{1}{s}\left(
    1 - e^{- \sqrt{1/[\widetilde{\eta}(s)D]}x_0}
    \right).
    \label{eq:S0-general}
\end{equation}
Using Eq. (\ref{mean under restart}) the mean first passage time becomes
\begin{equation}
\langle T_{r}\rangle = 
\frac{1}{r} \left(e^{\sqrt{1/[\widetilde{\eta}(r)D]}x_0}-1\right),
\label{Eq:MFATFiniteSub-general}
\end{equation}
which is valid for various choices of the kernel $\eta(t)$. 
As an illustrative case, we consider the ordinary subdiffusion \cite{maso2019transport,masoliver2019anomalous,kusmierz2019subdiffusive} for which $\eta(t)=\frac{t^{\mu-1}}{\Gamma(\mu)}$ so that $\widetilde{\eta}(s) \sim s^{-\mu}$ with $0 <\mu \leq 1$, and the waiting time distribution between the jumps has a power-law decay $\phi(t)\sim t^{-1-\mu}$. Following Eq.~(\ref{eq:S0-general}), we find
\begin{equation}
    \widetilde{Q_0}(s|x_0) =  
    \frac{1}{s}\left(
    1 - e^{- \sqrt{s^{\mu}/D}x_0}
    \right), 
    \label{eq:S0}
\end{equation}
and the corresponding MFPT reads
\begin{equation}
\langle T_{r}\rangle = 
\frac{1}{r} \left(e^{\sqrt{{r^{\mu}}/{D}}|x_0|}-1\right).
\label{Eq:MFATFiniteSub}
\end{equation}
Equation (\ref{Eq:MFATFiniteSub}) is consistent with a more general formula 
derived with different methods in \cite{shkilev2017continuous}. For $r\rightarrow0$ and $r\rightarrow\infty$, $\langle T_{r}\rangle$ diverges and in between there is a minimum for an optimal $r^*$ 
which can be found from the following transcendental equation \cite{stanislavsky2021optimal}
\begin{align}
    1-e^{-\zeta}=r\frac{d\zeta}{dr}, \quad \zeta=|x_0|/\sqrt{\widetilde{\eta}(r)D},
    \label{ORR-sub}
\end{align}
Eq. (\ref{ORR-sub}) is, in principle, solvable for a given waiting time kernel. In particular, for subdiffusion case, Eq. (\ref{ORR-sub}) boils down to 
$1-e^{-\zeta}=\frac{\mu \zeta}{2}$. The case $\mu=1$ reproduces the original diffusion, but otherwise renders a unique solution for the optimal scaled resetting rate $\zeta$ in terms of the subdiffusive index $\mu$ \cite{stanislavsky2021optimal}. Since for a subdiffusive particle, the MFPT to reach the origin (in the absence of
resetting) is infinite, it is only natural that resetting will stabilize the system rendering a finite MFPT and consequently, an optimal resetting rate.

\begin{figure}[t]
\centering
\includegraphics[width=8cm]{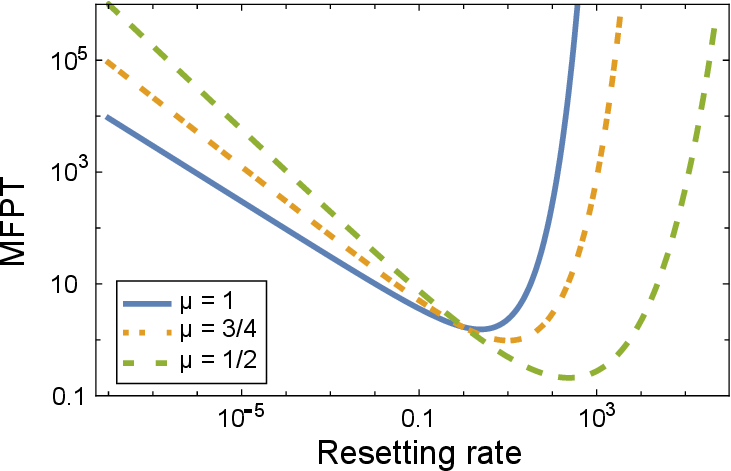}
\caption{\textbf{Mean first passage time for general diffusive search process under resetting}. MFPT for normal diffusion under stochastic resetting (\ref{mfpt_Br}) with $\mu=1$ (blue solid line), and MFPT for anomalous diffusion under stochastic resetting (\ref{Eq:MFATFiniteSub}) with $\mu=3/4$ (red dotted line), and $\mu=1/2$ (green dashed line). The parameters are set as: $D=1$, $x_0=1$. In all the cases, we observe an expedited completion due to resetting.}
\label{Fig2}
\end{figure}

\subsection{Heterogeneous diffusion with stochastic resetting}
Consider independent random walkers on a one dimensional heterogeneous medium, initially located at position $x_0$ and then they search for a target which is located at the origin. The target position
defines a bound of the search domain and the process is completed as soon as the target is first
detected. The evolution for the probability density function in this case can be written as \cite{dos2022efficiency,sandev2022heterogeneous}
\begin{align}
    \frac{\partial P(x,t)}{\partial t}=\frac{\partial^2}{\partial x^2} \left[ D(x)P(x,t) \right]+(A/2-1) \frac{\partial}{\partial x} \left[ D'(x)P(x,t) \right] ,
\end{align}
where $D(x)>0$ is a space-dependent diffusion coefficient manifesting for the heterogeneous medium and $0 \leq A \leq 2$ is a parameter that takes different values depending on the numerical scheme. For instance, the case of $A = 2$ (It\^{o} convention) is commonly used in finance \cite{ito1944109}, while $A = 1$ (Stratonovich convention) is popular in physics \cite{stratonovich1966new}. The highly anticipating case $A = 0$ (known as isothermal, kinetic, or H\"{a}nggi-Klimontovich) also has applications related to
Fick’s law \cite{hanggi1978stochastic,klimontovich1990ito,cherstvy2013anomalous}.  To mimic the heterogeneous diffusivity, it is often considered $D(x)=D |x|^\nu$ ($\nu<2$) which has been used extensively
to capture the diffusive motion of a particle on fractal objects and diffusion in turbulent media. For different interpretations of the heterogeneous diffusion processes we refer to \cite{ito1944109,stratonovich1966new,hanggi1978stochastic,klimontovich1990ito,cherstvy2013anomalous,leibovich2019infinite,sandev2022heterogeneous}. For this set-up, the survival probability in Laplace space reads~\cite{dos2022efficiency}
\begin{align}
    \widetilde{Q_0}(s|x_0)= \frac{1}{s}-\frac{2}{\Gamma(b)s} \left[ \frac{x_0^{\frac{2-\nu}{2}}}{2-\nu} \sqrt{\frac{s}{D}} \right]^b \times K_b\left( \frac{2 x_0^{\frac{2-\nu}{2}}}{2-\nu} \sqrt{\frac{s}{D}} \right),
\end{align}
where $K_b(z)$ is the modified Bessel function and $b=\frac{1-\nu(1-A/2)}{2-\nu} \geq 0$. The MFPT under resetting then can be obtained using Eq. (\ref{mean under restart})
\begin{align}
    \langle T_r \rangle=\frac{1}{r} \left[  \frac{\Gamma(b) \left( \frac{x_0^{\frac{2-\nu}{2}}}{2-\nu} \sqrt{\frac{r}{D}} \right)^{-b}}{2  K_b\left( \frac{2 x_0^{\frac{2-\nu}{2}}}{2-\nu} \sqrt{\frac{r}{D}} \right)} -1\right],
\end{align}
Recalling the definition of the Gamma function, we observe from the above that the MFPT is finite for $b>0$. For the resetting free process, the underlying first passage time density has an asymptotic tail $t^{-b-1}$ which is a generalized version of the simple diffusion case. In such cases, underlying process may have diverging MFPT depending on $\nu$ and $A$. Similar to the cases in the above (eg, diffusion), one can explicitly show that the MFPT becomes finite in the presence of resetting \cite{dos2022efficiency,ray2020space}. 

\section{When resetting works?}
\label{sec:when-resetting-works}
There are endless variety of ways in which first passage processes and resetting mechanisms mix and match. Rigorous studies show as is also evident from the examples above that for the processes like diffusion or subdiffusion, resetting always expedites the search while for the search processes in a confined domain or in the presence of a generic potential landscape, resetting can often be detrimental. A lot of efforts has been given in finding the physical conditions which marks this behavioral transition. To see this, one usually looks into a generic first passage process (with well defined first and second moments), add an infinitesimal resetting rate and try to examine its ramifications. Recalling the MFPT under resetting from Eq.~(\ref{mean under restart}) and expand it in the following polynomial in the small $r$ limit \cite{pal2017first,pal2019landau}
\begin{align}
\langle T_{ r} \rangle=a_0+a_1 r+a_2 r^2+a_3 r^3+ \cdots~,
\label{L1}
\end{align}
where $a_i$-s are the expansion coefficients.
Such kind of expansion respects a set of postulates which we state below. By taking $r$ strictly to be zero, we see that $\langle T_r \rangle=a_0=\langle T \rangle$, the MFPT  of the underlying process in the absence of resetting and is assumed to be positive finite.\\

\noindent
\textit{Physical meaning of the coefficients}.--- 
The coefficients $a_i$-s, so far defined formally, can be given physical meaning by analyzing $\langle T_{ r} \rangle$ around $r=0$. Recalling $\widetilde{f_T}(r|\cdot)=\int_0^\infty~dt~e^{-rt}~f_T(t|\cdot)$ to be the Laplace transformation of the underlying first passage time, expanding the same in the small $r$ limit in Eq. (\ref{mean under restart}) and furthermore comparing with the Taylor's series expansion of $\langle T_{ r} \rangle$ in Eq. \ref{L1}, we identify $a_0=\langle T \rangle,~a_1= -\frac{\langle T^2 \rangle}{2} + \langle T \rangle^2,~a_2= \frac{1}{6} \langle T^3 \rangle +\langle T \rangle^3-\langle T \rangle\langle T^2 \rangle,~a_3=-\frac{\langle T^4 \rangle}{4!} +\frac{\langle T^3 \rangle\langle T \rangle}{3}+\frac{\langle T^2 \rangle^2}{4}-\frac{3\langle T^2 \rangle\langle T \rangle^2}{2}+\langle T \rangle^4$, and so on where $\langle T^n \rangle$ is the $n$-th moment of the underlying first passage time distribution \cite{pal2017first,pal2019landau}. The moments (hence the coefficients) are explicit functions of the system parameters, and will characterize the transitions. \\

\begin{figure}[t]
\centering
\includegraphics[width=\linewidth]{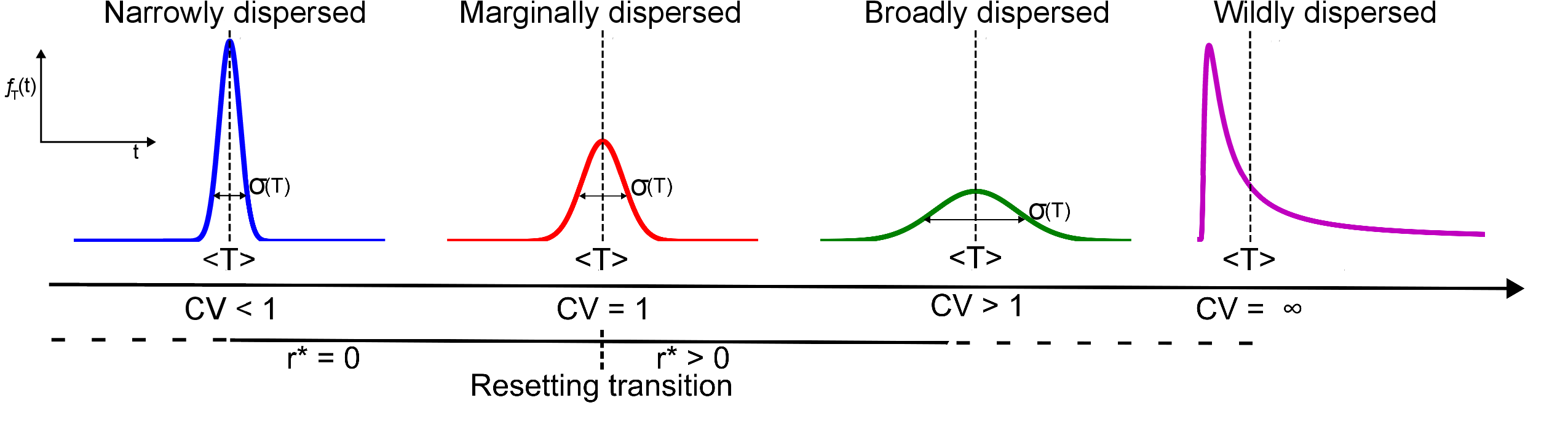}
\caption{Schematic illustration of the possible roles of resetting on the underlying first passage time processes. For the narrowly dispersed distribution, $CV<1$ and resetting is only detrimental. As $CV$ increases and crosses the boundary $CV=1$, the effect of resetting becomes more significant. Infinite dispersion $CV=\infty$ may arise when the underlying first passage time process is characterized by heavy power law tails (namely $f_T(t) \sim t^{-(1+\gamma)}$ with $0<\gamma<2$). Remarkably, in such cases one always observes an overarching gain due to resetting. The interpolation between these behaviors is captured by the optimal resetting rate $r^*$ which plays the role of an order parameter. The transition between $r^*>0$ and $r^*=0$ at the phase-separatrix $CV=1$ is often known as the \textit{resetting transition}.}
\label{resetting-transition}
\end{figure}

\noindent
\textit{A universal criterion.---}
For resetting to reduce the underlying MFPT, one would naturally expect $\langle T_{ r \to 0} \rangle < \langle T_{ r=0} \rangle$ with the introduction of resetting. This essentially means from Eq. (\ref{L1}) that in the linear order expansion one must have $a_1<0$. Rearranging $a_1$ from the above, one finds \cite{pal2017first,pal2019firstbranch}
\begin{align}
    CV \equiv \frac{\sigma(T)}{\langle T_{ } \rangle}>1
    \label{CV-criterion}
\end{align}
where $CV$ is called the the coefficient of variation.  This is a measure of statistical dispersion that stands for the ratio between the standard deviation $\sigma(T)$ and the mean $\langle T \rangle$ of the underlying first passage time $T$. The criterion is completely universal, system independent and essentially tells that for resetting to be beneficial this measure has to be higher than unity. Note that the $CV$-criterion depends only on the first two statistical metrics (which need to be finite) of the underlying process and not the entire distribution. For the stochastic processes, where $CV=\infty$ (broad distribution/heavy tail) or even not well-defined (eg., L\'evy first passage time distribution for the 1D simple diffusive search), resetting is guaranteed to help and thus this relation becomes redundant. Clearly, for the underlying first passage processes with narrow distributions (towards a deterministic limit) such that $CV<1$, resetting can only prolong the completion (see Fig. (\ref{resetting-transition})). It is worth noting that the $CV=1$ condition can also be derived from the \textit{inspection paradox} in probability theory \cite{pal2022inspection}. 

While the $CV$-criterion is the sufficient condition, it is not a necessary one. As seen above, the sign of $a_1$ is important to derive this criterion, whereas $a_2$ plays the pivotal role for higher order corrections. In such cases, $CV$-criterion need not be respected yet resetting can be found to be useful \cite{pal2019landau,ahmad2023comparing}.\\

\noindent
\textit{Interpretation of the criterion.---} The coefficient of variation $(CV)$ usually characterizes how broad or narrow a distribution is around its mean and that certainly depends on the underlying search mechanism, search domain, target configurations and many other intrinsic parameters. Recent single molecule experiments with colloidal particles, biomolecules, enzymes have brought deep insights into the first passage phenomena while looking into various microscopic search mechanism such as surmounting activation barriers, bottlenecks, gated reactions, transport across channels, facilitated diffusion and chemical reactions (see eg \cite{satija2020broad,sturzenegger2018transition,thorneywork2020direct} and  \cite{metzler2014first} for a review). Many such transport, escape or activation processes take place in multidimensional energy landscape with metastable states in the presence of deep kinetic traps and large fluctuating barriers. There, the microscopic searchers tend to spend exceedingly large time and thus the completion time statistics render to heavy tail broad distributions \cite{klafter2011first,bouchaud1990anomalous}. Similar arguments also apply to chemical reactions which are inherently stochastic in nature. Thus the outcome of the reactions are also random -- this means various kinds of products can be formed from an enzyme due to the intrinsic fluctuations. Enzymatic catalysis process often can take place on-pathways or off/parallel-pathways \cite{moffitt2014extracting,english2006ever}. In the former case, the reactions are usually nearest neighbour and the dwell time distribution is narrower. The latter case however can contain multiple pathways with active catalytic states with relative weights. This gives rise to multiple class of dwell time distributions which have long, multi-exponential or heavy tails. Thus the natural reaction time will have broader distributions with $CV>1$. These kind of scenarios also take place when searchers are macroscopic eg. foraging animals, falcons or even drones \cite{viswanathan2011physics,viswanathan1999optimizing}. Many such processes take place in unconfined territory in search of a steady supply of nutrients and other essential resources. Indeed, there can be large fluctuations around the average search time due to uncertainty, lack of cognition and experience and these can turn out to be deleterious and may often result in death. As shown above, resetting/quenching to the initial configuration for the microscopic searchers, unbinding of the enzymes from the metastable states or returning to the home for macroscopic searchers can reverse the deleterious effects of large fluctuations in the underlying search time, thus turning a marked drawback into a favorable advantage. Simply put, such \textit{reset kind events} can curtail the long detrimental trajectories and make the process more regular and remarkably efficient. From a broader perspective, whenever we have $CV>1$ (ie, worse search conditions for the underlying process), physical scenarios naturally suggest that search with resets/returns may be considered as a useful bet-hedging strategy. \\

\noindent
\textit{Optimal resetting rate as an order parameter and resetting transition.---} The equality $CV=1$ serves as a sharp boundary for resetting transition. This is not really a thermodynamic phase transition, but mimics the behavioral transition between the phases namely ``resetting-detrimental'' and ``resetting-beneficial''. Assume $CV<1$ - in this case, $\langle T_{ r \to 0} \rangle > \langle T_{ r=0} \rangle$ and thus the optimal resetting rate $r^*$ that globally minimizes $\langle T_r \rangle$ is fixed at zero. In other words, resetting only delays the completion process. On the other hand, for $CV>1$, we have $\langle T_{ r \to 0} \rangle < \langle T_{ r=0} \rangle$ and thus the optimal resetting rate $r^*$ that globally minimizes $\langle T_r \rangle$ takes a non-zero value. Simply put, an optimally reset process can always expedite the completion making resetting beneficial. The transition between $r^*=0$ to $r^*>0$ as we vary $CV$ is known as the ``resetting transition'' and the optimal resetting rate can be regarded as the \textit{order parameter} for such transition in resetting systems (see Fig.~(\ref{resetting-transition})). Since $CV$ is usually a function of the system parameters, one could imagine that there is a control parameter of interest, say $p$, that can be varied to span from $CV<1$ to $CV>1$. Henceforth, $CV=1$ would render a critical $p_c$ in the parameter space where the exact resetting transition will take place. In fact, one can show that near the transition, $r^* \sim |p-p_c|^\beta$, where $\beta=1$ is a critical exponent that is found to be universal across different reset processes~\cite{pal2019landau,ahmad2019first}. This is somewhat reminiscent of the continuous phase transition in statistical physics. Furthermore, a Landau like mean field approach can be used to probe such universality~\cite{pal2019landau}. \\

\section{Applications of resetting to real world search problems}
\label{sec:applications}

So far, we discussed the theoretical implications of stochastic resetting on search problems. In this section, we shift our perspective and turn our attention to the applications of stochastic resetting in real-world search problems. In particular, we discuss recent developments on applying resetting to home range search~\cite{pal2020search,tal2020experimental}, resetting facilitated diffusive transport~\cite{jain2023fick}, turnover of chemical reactions~\cite{reuveni2014role,rotbart2015michaelis,biswas2023rate}, and income dynamics~\cite{jolakoski2023first}. This helps us bridge the gap between theoretical research and its tangible applications in real-world scenarios.

\subsection{Home range search}
One of the intriguing applications of random resetting in search problems is the home range search. In this context, a searcher is not indefinitely adrift in a search space but has a reference point — a ``home'' — to which it can return under certain conditions. This kind of search strategy finds applications in animal foraging behavior, robot exploration, and many areas where search optimization is crucial.

Imagine a searcher that begins at an origin or ``home'' in a potentially $d$-dimensional search space $\mathcal{D}$. This searcher attempts to locate a target (or multiple targets). For the unconstrained search, the searcher might find the target in a random time $T$, taken from the distribution $f_T(t)$. However, real-world searchers often have limitations or strategies that cause them to return to a starting or known point. In our model, if the searcher does not find the target within a stipulated time $R$ distributed according to $f_R(t)=re^{-rt}$, it retreats back to its home. The time it takes to return, denoted as $\tau(\vec{x})$, can vary depending on where the searcher is when the \textit{decision to return home} is made. Once the searcher reaches home, it might rest or re-strategize, staying there for a time span $W$. Post this, the search initiates anew, making the process cyclic \cite{pal2020search}.

To make this model tractable, we employ two assumptions. First, targets cannot be located during the phases of return and waiting at home. Second, search cycles are isolated events. This means each new search starts with no memory of past endeavors, thus ensuring independence between cycles. The latter fact allows one to utilize the renewal approach as sketched out in Sec \ref{sec:formalism} and estimate the MFPT for such home-range-search process which reads \cite{pal2020search}
\begin{align}
\langle T_r  \rangle 
=\underbrace{ \frac{\widetilde{Q_0}(r|\Vec{x_0})}{1-r \widetilde{Q_0}(r|\Vec{x_0})}}_{\text{search}}+\underbrace{ \frac{r \int_{\mathcal{D}}d\Vec{x}~ \tau(\Vec{x}) \widetilde{P}_0(\vec{x},r)}{1-r \widetilde{Q_0}(r|\Vec{x_0})} }_{\text{return}}+\underbrace{ \frac{r\widetilde{Q_0}(r|\Vec{x_0})}{1-r \widetilde{Q_0}(r|\Vec{x_0})} \langle W \rangle}_{\text{home}} ,
\label{MFPT-HRS}
\end{align}
where $ \tau(\vec{x})$ is the home-return time of the searcher starting from the coordinate $\Vec{x}$ at the time of resetting. Here, $\widetilde{P}_0(\vec{x},r)=\int_0^\infty dt ~e^{-rt}P_0(\Vec{x},t)$ is the LT of $P_0(\Vec{x},t)$ -- PDF of the underlying process such that $Q_0(t|\Vec{x_0})=\int_{\mathcal{D}}~d\Vec{x}~ P_0(\Vec{x},t)$ is the survival probability of the searcher upto time $t$.

\subsubsection{Return of a diffusive searcher with a constant velocity}
\begin{figure}[t]
    \centering
    \includegraphics[width=\linewidth]{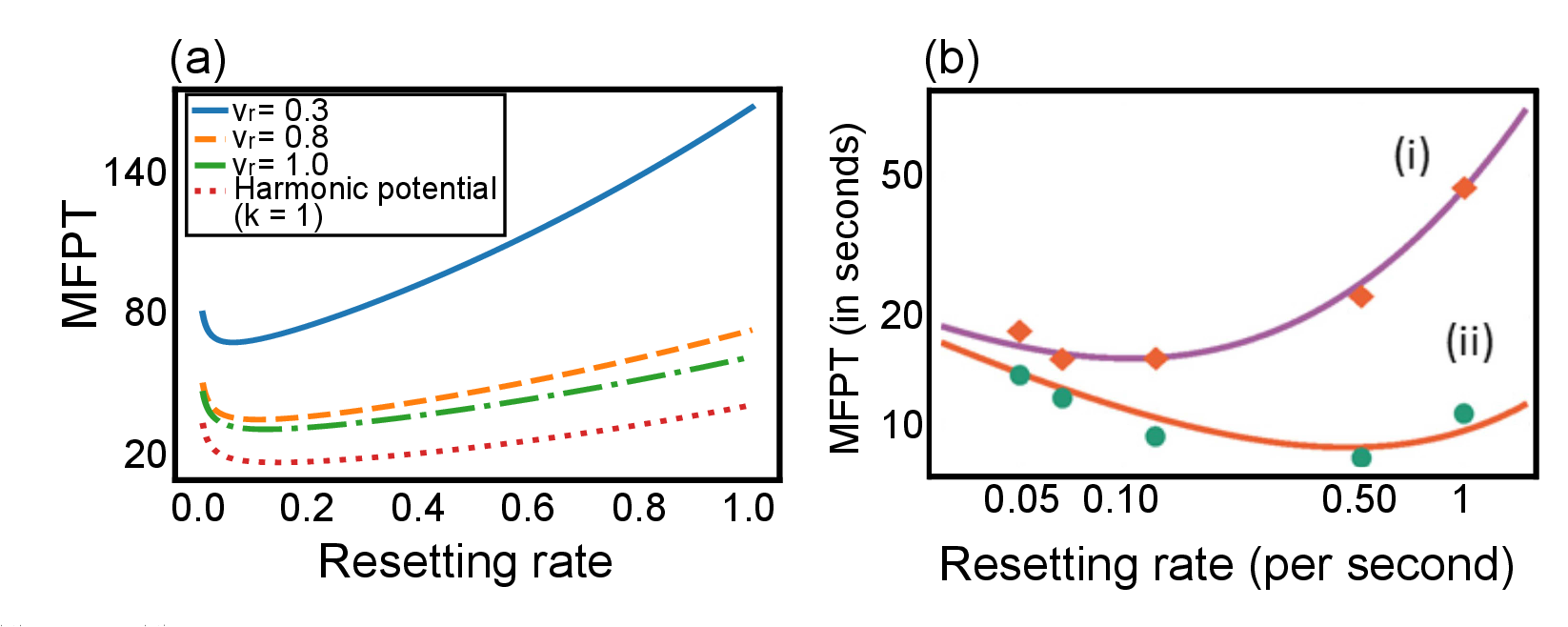}
    \caption{\textbf{Behaviour of the mean first passage time of a home range diffusive searcher as a function of resetting rate under various modes of return}. Panel \textbf{(a)} shows MFPTs when the searcher returns to its home at a constant velocity or is attracted by a harmonic trap centered at its home. Panel \textbf{(b)} shows MFPTs of a silica particle to a target wall under non-instantaneous returns with (i) $\tau_0 = 3.79$s and instantaneous returns with (ii) $\tau_0 = 0$ respectively. Theoretical predictions (Eq. (\ref{return-const-time}), solid lines) are in good agreement with experimental data (symbols). Panel \textbf{(b)} is adapted from \cite{tal2020experimental}.}
    \label{fig:home-range-search}
\end{figure}

To exemplify Eq. (\ref{MFPT-HRS}), let us consider a one dimensional diffusive search process where the searcher has to find a target placed at $L$. The searcher starts from its home which is fixed at the origin. Let us assume that upon resetting, the searcher returns to its home with a constant velcity $v_r$ such that $\tau(x)=|x|/v_r$. In this case, elementary calculation yields $\widetilde{P}_0(\vec{x},r)=\frac{1}{\sqrt{4Dr}} \left[ e^{-\sqrt{r/D}|x|} -e^{-\sqrt{r/D}(2L-x)} \right]$ and thus from Eq. (\ref{MFPT-HRS}) one finds \cite{pal2020search}
\begin{align}
    \langle T_r \rangle = \frac{1}{r} \left( e^{\sqrt{r/D}L}-1 \right)+ \frac{L}{v_r} \left( \frac{2 \sinh{(\sqrt{r/D}L)}}{\sqrt{r/D}L} -1 \right),
\end{align}
where we have assumed $W=0$ for simplicity. Note that the first term on the RHS is accounted for the search process and thus identical to the MFPT under instantaneous resetting. The second term on the RHS accounts for the delay due to the finite time return to its home. As $v_r \to \infty$, one recovers the instantaneous limit since the searcher returns to its home almost instantaneously. See Fig. \ref{fig:home-range-search}a which shows the modulation of MFPT under different return velocity.

\subsubsection{Return of a searcher under the action of a harmonic potential}
We now consider a return process where a generic one dimensional searcher returns under the influence of a harmonic potential $U(x)=\frac{1}{2}kx^2$ such that we can write a Newton's law for the return process $\Ddot{x}=-kx$. This is a deterministic return process and thus the solution of the return path is $x=x_0 \cos{(\sqrt{k}t)}$, where $x_0$ is the starting location. Using this one finds the return time to the origin to be $\tau(x)=\pi/\sqrt{4k}$, which turns out to be independent of $x$. Plugging this in Eq. (\ref{MFPT-HRS}), one finds
\begin{align}
    \langle T_r \rangle 
= \frac{\widetilde{Q_0}(r|x_0)}{1-r \widetilde{Q_0}(r|x_0)}+ \frac{\pi}{\sqrt{4k}} \frac{r\widetilde{Q_0}(r|x_0)}{1-r \widetilde{Q_0}(r|x_0)} ,
\end{align}
which holds for any arbitrary underlying search process.
In particular, for diffusive process one has \cite{bodrova2020resetting}
\begin{align}
    \langle T_r \rangle 
=  \left[\frac{1}{r}+  \frac{\pi}{\sqrt{4k}}\right] \left( e^{\sqrt{r/D}L}-1 \right) .
\end{align}
When the stiffness of the potential is very high (ie, $k \to \infty$), one immediately recovers the instantaneous return limit. Otherwise for finite stiffness, it can often facilitate the search compared to the velocity driven return process (see Fig. \ref{fig:home-range-search}a).

\subsubsection{Return in first passage time under fixed time -- resetting experiments}
We now turn our attention to the case when the searcher returns to its home in fixed time say $\tau_0$. This protocol is quite feasible in the single particle experiments since the searcher (say, a silica particle) can be manipulated with great precision by optical traps \cite{tal2020experimental}. In such case, the MFPT reads
\begin{align}
\langle T_r  \rangle 
= \frac{\widetilde{Q_0}(r|x_0)}{1-r \widetilde{Q_0}(r|x_0)}+ \tau_0 \frac{r\widetilde{Q_0}(r|x_0)}{1-r \widetilde{Q_0}(r|x_0)} ,
\end{align}
which again holds for any arbitrary underlying search process. In particular, in the first passage time experiment \cite{tal2020experimental}], resetting mechanism was conducted stochastically using holographic optical tweezers on a silica particle (which mimics a diffusive dynamics). Following the resetting, the particle was returned to its home (ie, the initial coordinate) using the optical traps in a constant time $\tau_0$ \cite{tal2020experimental}. For a diffusive particle starting from the origin, the above expression for the MFPT to a target set at $L$ simplifies to
\begin{align}
    \langle T_r \rangle = \left[ \frac{1}{r}+\tau_0  \right] \left(e^{\sqrt{r/D}L} -1\right),
    \label{return-const-time}
\end{align}
which was verified against the experimental data to find an excellent fit \cite{tal2020experimental}. See Fig. \ref{fig:home-range-search}b which was adapted from \cite{tal2020experimental}.

\subsubsection{Stochastic return of the searcher}
Notably, we have assumed that the return to the home is a deterministic process making $\tau(\Vec{x})$ a deterministic function once $\Vec{x}$ is fixed. One can also consider a situation where the return processes are stochastic. This is natural in many set-ups since there will always be random fluctuations that are uncontrollable and thus return protocols need not be absolutely deterministic. In such cases, $\tau(\Vec{x})$ becomes a stochastic function of the return path and thus to emphasize this additional averaging, one needs to replace $\tau(\Vec{x})$ by $\overline{\tau(\Vec{x})}$ in Eq. (\ref{MFPT-HRS}). Interested readers can take a look at these references \cite{gupta2020stochastic,gupta2021resetting,biswas2023stochasticity,mercado2020intermittent} for such scenarios.

\subsection{Resetting facilitated diffusive transport through channels}
Diffusion of particles, molecules, or even living microorganisms in confined geometries such as tubes and channels plays a key role across various scales in natural and technological processes \cite{jacobs1935diffusion,zwanzig1992diffusion}. A diffusing particle inside a channel can either escape from both sides of the membrane or it may be allowed to escape through one side while the other side simply reflects it back. Key quantities in these processes are the first passage probabilities and average escape times from the channel (or lifetimes) conditioned to the exit side of the membrane, as well as the overall average lifetime of the particle in the channel \cite{berezhkovskii2011time,berezhkovskii2006identity}. These quantities are ubiquitous in theoretical studies on channel-facilitated transport. Enormous theoretical efforts have been devoted over the years to make the transport inside various channels more efficient. The fact that resetting has the ability to speed-up complex search processes naturally compels one to use resetting as a useful strategy for the  channel-facilitated first passage transport. Indeed this was shown in \cite{jain2023fick} recently, and we review the results in brief. 

The canonical Fick-Jacobs formalism is a key approach to treat diffusion of particles, ions, molecules, or even living microorganisms in confined geometries such as tubes and channels of varying cross-sections \cite{jacobs1935diffusion,zwanzig1992diffusion,dagdug2003diffusion}. For instance, the effective one-dimensional Fick-Jacobs equation for a diffusing particle, starting at $x_0$, inside a three dimensional conical tube of variable radius $R(x)$ and length $L$ can be written as
\begin{equation}
\frac{\partial {P}(x,t)}{\partial t}=\frac{\partial}{\partial x}\left\{D(x)e^{-\beta_\mathcal{T} U(x)}\frac{\partial}{\partial x}\left[e^{\beta_\mathcal{T} U(x)} P(x,t) \right] \right\},
\label{Sm}
\end{equation}
where the entropy potential is given by -$\beta_\mathcal{T} U(x)=\ln(R^2(x)/R^2(x_0))$ for a 3D tube. Here, $\beta_\mathcal{T} = 1/(k_B \mathcal{T})$ is the inverse temperature with $k_B$ the Boltzmann constant. Replacing $U(x)$ for the 3D cone, the modified equation reads
\begin{align}
\frac{\partial {P}(x,t)}{\partial t}=&  D_\lambda \frac{\partial}{\partial x}\left\{ R^2(x)\frac{\partial}{\partial x} \left[\frac{{P}(x,t)}{R^2(x)} \right]\right\}, \label{propagator}
\end{align}
where $D_\lambda$ is given by the Reguera-Rubi 
formula 
$D_\lambda=\frac{D_0}{\sqrt{1+\lambda^2}}    
$ \cite{reguera2001kinetic}.
We further assume that the tube radius $R(x)$ increases in the axial direction with a constant rate $\lambda$ so that $R(x)=a+\lambda x$, where the $x$-coordinate is measured along the tube axis, and $a$ is the tube radius at $x=0$. 
For a conical tube
with two absorbing points at $x=0$ and $x=L$,  the unconditional first passage time density (in Laplace space) is given  by \cite{jain2023fick}
\begin{align}\label{eq:fsaa}
   \widetilde{f_T}(s|x_0)=\frac{(\lambda L+a)\sinh( \sqrt{s/D_\lambda} x_{0})+a \sinh(\sqrt{s/D_\lambda} (L-x_{0}))}{(\lambda x_{0}+a)\sinh(\sqrt{s/D_\lambda}L)}.
\end{align} 
The first passage density can then be substituted in Eq. (\ref{mean under restart}) to obtain the MFPT of the searcher in the presence of resetting \cite{jain2023fick}
\begin{align}\label{T2abs}
    \langle T_{r}(x_{0}) \rangle=\frac{(\lambda x_{0}+a)\sinh(\alpha_\lambda L)-a \sinh(\alpha_\lambda (L-x_{0}))-(\lambda L+a)\sinh(\alpha_\lambda x_{0})}{r[a \sinh(\alpha_\lambda (L-x_{0}))+(\lambda L+a)\sinh(\alpha_\lambda x_{0})]},
\end{align}
where $\alpha_\lambda=\sqrt{r/D_\lambda}$. It can further be shown using the $CV$-criterion ( \ref{CV-criterion}) that resetting is going to be beneficial (i.e., an optimal resetting rate $r^*>0$ would exist) when the following condition is satisfied \cite{jain2023fick}
\begin{align}\label{eq:critaa}
    (\lambda v+\tilde{a})
    \left[\lambda(1+v)(7-3v^2)+15\tilde{a}(1+v-v^2) \right] 
    \geq 10v(1-v)
    \left[\lambda(1+v)+3\tilde{a} \right]^2,
\end{align}
where $\tilde{a}=a/L$ and $v=x_{0}/L$. Thus, the criterion is fulfilled as long as the particle starts sufficiently close to one of the absorbing boundaries ($v\sim 0$ or $v\sim 1$) but not when it starts out in the center. For starting positions near the center of the tube, increasing the reset rate increases the MFPT since any trajectory on both sides is going to take particle closer to the boundary and reset hinders it. But for starting positions closer to the boundaries, resetting decreases the lifetime because there are now many more possible trajectories that are taking the particle away from the boundary and resetting systematically eliminates them \cite{jain2023fick}. A carefully navigated resetting strategy thus can facilitate\textit{ diffusive transport} through narrow channels. Likewise, resetting can also be useful for \textit{active transport} and search in a confined arena \cite{sar2023resetting}.

\begin{figure}[t]
    \centering
    \includegraphics[width=8.7cm]{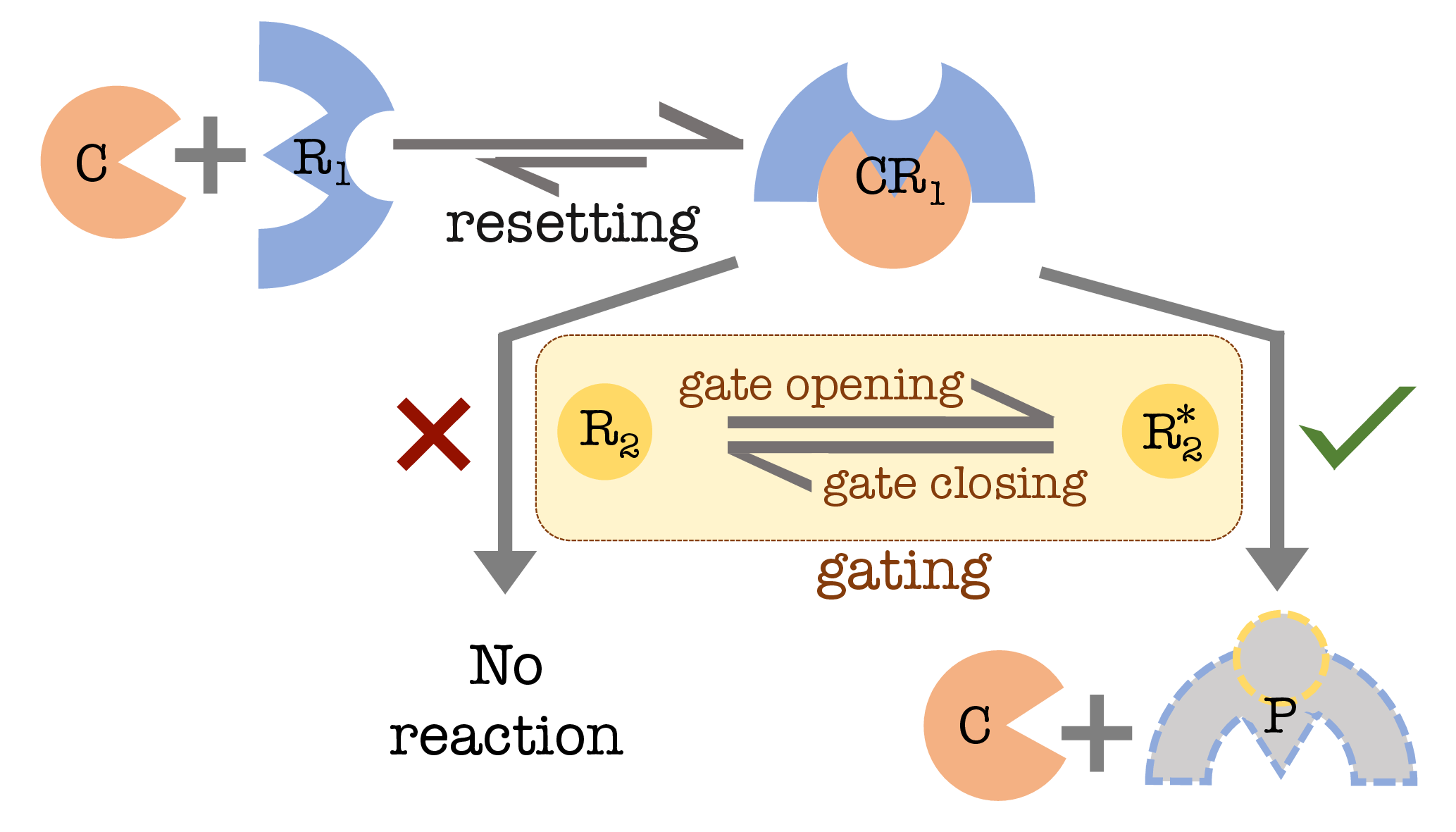}
    \caption{Scheme for a gated chemical reaction between two reactants $R_1$ and $R_2$, catalysed by $C$. Initially, the catalyst binds reversibly with $R_1$ to generate $CR_1$, a metastable intermediate: $C+R_1\stackrel{}{\rightleftharpoons}CR_1$. In the next step, $CR_1$ reacts selectively with  $R^{\star}_2$ (the {\it reactive} or {\it open-gate} state of $R_2$) to generate the product $P$ and liberate the catalyst: $CR_1+R_2^{\star}\to C+P$. This step can be modeled by gated drift-diffusion, while the unbinding of $C$ from $R_1$ is essentially manifested as resetting. Adapted from \cite{biswas2023rate}.}
    \label{Fig1-chem}
\end{figure}

\subsection{Turnover of the gated chemical reactions}
For a reaction, two molecules first need to meet by overcoming the activation barrier between them. Since the motion of the molecules are stochastic, these reactions are inherently stochastic in nature and estimation of the reaction-time is often cast as a first passage problem \cite{english2006ever,iyer2016first,metzler2014first,moffitt2014extracting}. Chemical reactions usually comprise of three basic steps: binding (when one reactant $C$ binds another one $R_1$ to form a metastable complex $CR_1$), unbinding or resetting (when the complex $CR_1$ is disassociated to the parent reactants) and catalysis (when a product $P$ is successfully formed).  Often, the reactant molecules switch between a
reactive and a non-reactive state; and thus the collisions between the reactants must happen in their reactive states for the completion of a successful product formation (see Fig. \ref{Fig1-chem}). This is known as `gating' in biochemistry that typically refers to the transition
between the open and closed states of an ion channel; and the ions are
allowed to flow through the channel only when it is open \cite{bressloff2014stochastic,szabo1982stochastically}. 

Such gated reactions can be illustrated by
considering a diffusive transport process confined to the positive semi-infinite space in the presence of a target that randomly switches between an open and a closed state with constant rates. Consider that the target switches from the non-reactive to the reactive state at a rate $\gamma>0$, and the reverse transition takes place with a constant rate $\beta>0$. This is a general scenario mimicked by, e.g., a chemical reaction (see Fig. \ref{Fig1-chem}), where the collisions between reactants [$CR_1$ and $R_2$] lead to the formation of product only when at least one of the reactants is in an activated state [when $R_2$ exists as $R_2^{\star}$] and not otherwise. Since resetting is an integral part of any consecutive chemical reaction that has a reversible first-step [when $C$ binds to, or unbinds from, $R_1$], the reaction scheme shown in Fig. \ref{Fig1} can be modeled as gated diffusion process in a potential landscape with resetting. The mean turnover time of this reaction can then be obtained by casting into the formulation shown above \cite{biswas2023rate,mercado2021search}
\begin{align}
\langle T_{r}^G\rangle=\frac{e^{\sqrt{\frac{r}{D}}x_0}-1}{r}
+
\frac{\beta e^{\sqrt{\frac{r}{D}}x_0}}{\gamma \sqrt{r[r+\gamma+\beta]}}\mbox{\hspace{-0.15cm}}\left[1+\frac{r e^{-\sqrt{\frac{r+\gamma+\beta}{D}}x_0}}{(\gamma+\beta)}\right].
\label{mfpt_av_d}
\end{align}
It can be shown explicitly that resetting or unbinding is always going to expedite the turnover process compared to the underlying gated diffusive process \cite{biswas2023rate,mercado2021search}. 

\subsection{Resetting in income dynamics}

Income mobility describes the dynamic aspects of income within a society. Formally, it quantifies the time it takes for an individual to transition from one income status to another, shedding light on the pace and accessibility of upward mobility in different societies~\cite{JanttiJenkins2015,stojkoski2022measures,stojkoski2022ergodicity,la2023unraveling}. 

Recently, it was shown that a first passage under resetting approach to income dynamics can effectively capture micro-level variations and experiences of individuals, a caveat that is often missing in standard income mobility analyses. The idea is that a baseline model for income dynamics called geometric Brownian motion with stochastic resetting and allows us to develop a ``model'' based view for quantifying the time needed for an individual currently with income $x_0$ to reach \textit{target income} $y$~\cite{gabaix2016dynamics,nirei2004income,aoki2017zipf,stojkoski2022income,vinod2022nonergodicity,vinod2022time}.

\begin{figure}[t]
\centering
\includegraphics[width=\linewidth]{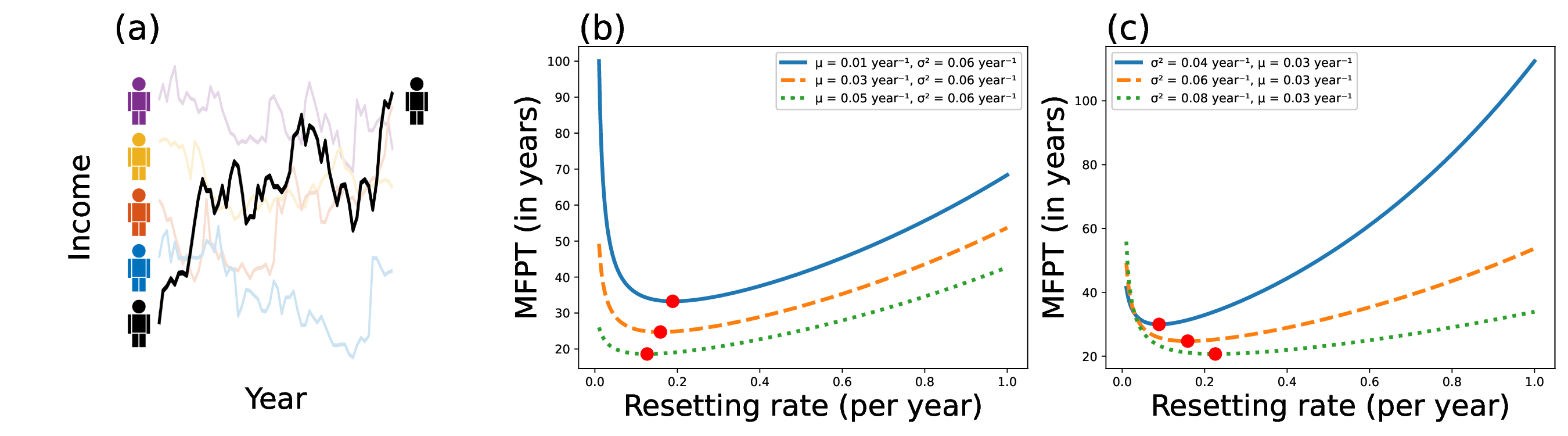}
\caption{\textbf{Resetting in income dynamics.} Panel \textbf{(a)}: A schematic figure for the income movement of an individual over the years. Panel \textbf{(b)}: MFPT (in years) as a function of the resetting rate for various growth rates $\mu$. Panel \textbf{(c)}: MFPT (in years) as a function of the resetting rate for various volatilities $\sigma^2$. For panels \textbf{(b-c)}: We set $x_0 = x_R = 1$ and $y = 2$. Adapted from~\cite{jolakoski2023first}.\label{fig:srgbm}}

\end{figure}

In this application, the particle position $x(t)$ can be thought of as the income of a worker in period $t$ (Fig.~\ref{fig:srgbm}(a)). This observable grows multiplicatively with a rate $\mu$ and volatility $\sigma$~\cite{cherstvy2017time,stojkoski2020generalised,stojkoski2019cooperation,kemp2022statistical,kemp2023learning} until a random event that occurs with a rate $r$ resets its dynamics~\cite{evans2011diffusion}. The reset event can be interpreted as a worker that left the job market (for example by retiring, being laid off, or after an injury) and is substituted by another younger worker with a starting income $x_R$ (here, $x_R$ need not be equal to $x_0$)~\cite{nirei2004income}. Hence, the MFPT in srGBM disaggregates the income dynamics to the level of individual workers, offering a more nuanced lens through which we can understand the intricacies of income mobility. 

To derive the MFPT in srGBM we use the Fokker-Planck equation for the distribution of the random income variable $x(t)$ that follows the geometric Brownian motion with $r=0$. This equation reads
\begin{align}
    \frac{\partial}{\partial t} P(x,t)=\mathcal{L}(x)P(x,t)~,
\end{align}
where $\mathcal{L}(z)=\mu z \frac{\partial}{\partial z}+\frac{\sigma^2}{2} z^2 \frac{\partial^2}{\partial z^2}$ is the generator for the GBM process (following It\^{o} convention)~\cite{stojkoski2020generalised,stojkoski2021geometric}. For Stratonovich and H\"{a}nggi-Klimontovich convention of the GBM we refer to \cite{sandev2020hitting,sandev2023special}. Using this equation and the general formalism for first passage, it can be shown that the Laplace form of the solution for the mean first passage time of a GBM trajectory to reach a finite income target $y$, starting at $x_0$, is
\begin{align}
   \widetilde{ f_T}(s|x_0)&=\left( \frac{x_0}{y} \right)^{q_1(s)}, \quad x_0 \leq y, \\
     q_1(s)&=\frac{\sqrt{(\sigma^2-2\mu)^2+8s \sigma^2}+(\sigma^2-2\mu)}{2 \sigma^2}>0.
   \label{MGF-LT}
\end{align}
Substituting the above in Eq. (\ref{mean under restart}), one obtains the MFPT of a srGBM trajectory that stochastically resets to $x_R$. 

Applying the srGBM framework to study the income dynamics yields enlightening observations. For instance, as the economy experiences a larger growth of income or larger volatility, the MFPT declines, as illustrated in Fig.~\ref{fig:srgbm}(b-c). This observation aligns with real-world insights: it is generally easier for workers to traverse the income distribution when the economy booms and there is increased randomness\cite{aristei2015drivers}.

The stochastic resetting rate $r$ in the srGBM framework provides another layer of depth to our understanding (depicted with red dots in Fig.~\ref{fig:srgbm}(b-c)). This rate, especially when optimized at $r^*$, can be perceived as a lever that policymakers might tweak to influence the rhythm of income dynamics within an economy. For instance, by calibrating policies around workforce participation - such as the frequency of retirements - authorities could potentially regulate the time it takes for a worker to navigate the income spectrum~\cite{cremer2004social,staubli2013does,sheather2021great}. Two salient features of the optimal resetting rate in srGBM emerge which mirror real-world economic observations. First, when volatility is held constant, a surge in growth ($\mu$) reduces the rate at which MFPT reaches its minimum. This, for example, could be a result of joint factors leading to economic growth such as efficient qualification programs, or quality foreign investments. Second, for a fixed growth rate, when the randomness in the system is increased, we also observe an increased optimal resetting rate. This, for example, can be a result of increased intrinsic differences between societal groups when it comes to getting new jobs (e.g., gender or racial differentials).

These theoretical insights have led to an empirical methodology for using the srGBM MFPT in real economies~\cite{jolakoski2023first,stojkoski2023income}. By integrating the srGBM framework with data, a more textured landscape of income dynamics and economic mobility emerges, paving the way for richer insights and more informed policy decisions. The simplicity of the srGBM MFPT approach paves the way for potential applications in other countries, opening doors to global economic insights.

\section{Discussion}
\label{sec:discussion}
In this chapter we have shown that stopping a complex search process intermittently only to reset or restart from scratch can often lead to an \textit{accelerated completion} -- a counter-intuitive effect that is orthogonal to our general perception. The subject of random resetting has been a topical interest and has seen a wide panorama of applications starting from physics, chemistry, biology, ecology, computer science to economics. Developing a comprehensive framework for the first passage statistics under resetting, we have applied the formalism to many diffusive and non-diffusive theoretical first passage models. In doing so, we discuss in detail the conditions which play a pivotal role in determining whether resetting can be indeed be used as a mitigating strategy. Finally, we showcase a few interdisciplinary applications of resetting motivated by the real world search processes. 

It will be quite interesting to study various non-Markovian complex search processes eg geometry-controlled kinetics \cite{benichou2010geometry}, random walks in fractal geometry \cite{condamin2007first} under generic resetting mechanism. Yet another interesting direction that remains quite less explored is the thermodynamical cost of first passage resetting (see \cite{pal2021thermodynamic,PhysRevE.108.044117,gupta2020work}). For instance, what will be the energy expenditure for an optimally reset process in comparison to a simply reset process? Will there be a universal thermodynamic trade-off relation at the optimality? Naturally, these questions are fundamental to design resetting strategies in living systems. Needless to say, with the current advent of the experimental studies using laser traps \cite{goerlich2023experimental,tal2020experimental,besga2020optimal}, tunable robots \cite{paramanick2023programming,altshuler2023environmental} or the applications of resetting in quantum search processes \cite{yin2023restart,kulkarni2023first}, queuing theory \cite{bonomo2022mitigating} \& material science \cite{aquino2023fluid}, the field marks the inception of a new era where we expect to see more realistic and trans-disciplinary applications of resetting in complex systems.

\begin{acknowledgement}
AP gratefully acknowledges
research support from DST-SERB Start-up Research Grant
No. SRG/2022/000080 and the DAE, Government of India. TS acknowledges financial support by the German Science Foundation (DFG, Grant number ME~1535/12-1). TS is supported by the
Alliance of International Science Organizations (Project No.~ANSO-CR-PP-2022-05). TS is also supported by the Alexander von Humboldt Foundation. We thank Arup Biswas for an illustration. 
\end{acknowledgement}
%
%
%

\appendix

\bibliographystyle{spphys.bst} 

\end{document}